\documentclass[onecolumn]{aastex6}

\usepackage{amsmath,amsfonts,amsthm}
\usepackage{amssymb}
\usepackage[T1]{fontenc}
\usepackage{graphicx}

\begin{document}
\title{Stochastic processes as the origin of the double-power law shape of the Quasar luminosity function}
\author{Keven Ren\altaffilmark{1,2}, Michele Trenti\altaffilmark{1,2}, Tiziana Di Matteo\altaffilmark{1,3}}
\affil{$^1$ School of Physics, The University of Melbourne, Parkville, Victoria, Australia \\
$^2$ ARC Centre of Excellence for All Sky Astrophysics in 3 Dimensions (ASTRO 3D) \\
$^3$ McWilliams Center for Cosmology, Department of Physics, Carnegie Mellon University, Pittsburgh, PA, 15213, USA \\
 }
\email{kevenr@student.unimelb.edu.au}

\begin{abstract}
The Quasar Luminosity Function (QLF) offers insight into the early co-evolution of black holes and galaxies. It has been characterized observationally up to redshift $z\sim6$ with clear evidence of a double power-law shape, in contrast to the Schechter-like form of the underlying dark-matter halo mass function. We investigate a physical origin for the difference in these distributions by considering the impact of stochasticity induced by the processes that determine the quasar luminosity for a given host halo and redshift. We employ a conditional luminosity function and construct the relation between median quasar magnitude versus halo mass $M_{UV,\rm{c}}(M_{\rm{h}})$ with log-normal in luminosity scatter $\Sigma$, and duty-cycle $\epsilon_{\rm{DC}}$, and focus on high redshift $z\gtrsim4$. We show that, in order to reproduce the observed QLF, the $\Sigma=0$ abundance matching requires all of the brightest quasars to be hosted in the rarest most massive dark-matter halos (with an increasing $M_{UV,\rm{c}}/M_{\rm{h}}$ in halo mass). Conversely, for $\Sigma>0$ the brightest quasars can be over-luminous outliers hosted in relatively common dark-matter halos. In this case, the median quasar magnitude versus halo mass relation, $M_{UV,\rm{c}}$, flattens at the high-end, as expected in self-regulated growth due to feedback. We sample the parameter space of $\Sigma$ and $\epsilon_{\rm{DC}}$ and show that $M_{UV,\rm{c}}$ flattens above $M_{\rm{h}}\sim 10^{12}M_{\odot}$ for $\epsilon_{\rm{DC}}<10^{-2}$. Models with $\epsilon_{\rm{DC}}\sim1$ instead require a high mass threshold close to $M_{\rm{h}}\gtrsim10^{13}M_{\odot}$. We investigate the impact of $\epsilon_{\rm{DC}}$ and $\Sigma$ on measurements of clustering and find there is no luminosity dependence on clustering for $\Sigma>0.3$, consistent with recent observations from Subaru HSC.
\end{abstract}

\section{INTRODUCTION} \label{sec:intro}

Quasars or quasi-stellar objects (QSOs) are highly luminous objects that are a class of active galactic nuclei (AGN) powered by the accretion of matter from a disk onto the central supermassive black hole (SMBH). While quasars have been observed at all redshifts up to $z \sim 7.5$ \citep{Ba_ados_2017}, inferences of SMBH masses from luminous high redshift quasars ($z > 6$) can reach up to $\sim 10^{9} M_{\odot}$. The progenitors of these extreme quasars, and the mechanism behind their immense rapid growth remain a topic of open active research and a stringent theoretical hurdle to overcome. At face value, the relative rarity of the luminous quasars in question (comoving densities of the order $\sim 10^{-9}$ Mpc$^{-3}$) would suggest that these objects reside at regions of extreme over-densities and contained within the most massive halos \citep{Springel_2005}. Recent work
however suggests this may not be the case \citep{Fanidakis_2013, Aversa_2015, Di_Matteo_2017} and that instead the first quasars may live in relatively more common halos.
Thus, studying the connection between the early quasars to their host halos can provide insight to the origin of the first quasars and their subsequent evolution over cosmic time.

The role of AGN in galaxy formation has been a topic of considerable research interest in the past decade. The remarkably tight local M-$\sigma$ relation hints towards a deep connection between a host galaxy and their respective black hole \citep{Gebhardt_2000,Kormendy_2013} and by extension, a connection between galaxies and their AGNs. The nature and the implications of this association have been explored both in observations \citep{Fabian_2012,Page_2012,Cicone_2014} and theory/simulations (\citealt{Di_Matteo_2003,Croton_2006, De_Lucia_2007, Sijacki_2007, Jahnke_2011, Conroy_2012}, EAGLE: \citealt{Schaye_2014}, Meraxes: \citealt{Qin_2017}, SAGE/RSAGE: \citealt{Croton_2016, Raouf_2017}, Horizon-AGN: \citealt{Beckmann_2017}, BLUETIDES: \citealt{Feng_2015, Ni_2018}). In this context, the luminosity functions of galaxies and AGNs are key observables that are easily measured owing to modern wide-area surveys such as 2dF Galaxy Redshift Survey (2dFGRS: \citealt{Colless_2001}), the Sloan Digital Sky Survey (SDSS, \citealt{Richards_2002}), and the Cosmic Assembly Near-infrared Deep Extragalactic Legacy Survey (CANDELS: \citealt{Grogin_2011, Koekemoer_2011}). On one hand, the galaxy luminosity function and that of its underlying dark-matter halo mass function (HMF) are well described by Schechter-like shapes (a power law with an exponential drop-off at the bright/massive end) to all redshifts \citep{Arnouts_2005,Alavi_2013,Weisz_2014,2015ApJ...803...34B, 2014ApJ...786...57S, Finkelstein_2015}, while the quasar luminosity function (QLF) has been canonically described at all redshifts by a double power law, or where faint end data is not available, by a single power law at its bright end  \citep{Croom_2009,Schneider_2010,Ross_2013, Akiyama2017, Kulkarni2019, Shen_2020}. Understanding why and how this functional shape transition happens from the distribution of halos/galaxies to luminous quasars would refine our understanding of the paradigm around halo-galaxy-AGN interactions. Furthermore, it would not only contribute to elucidate the subsequent evolution of the QLF, but also potentially enable additional constraints for simulations/theories on the formation and growth of galaxies and SMBHs.

In this work, we investigate a physical origin for the difference in the shapes of the QLF and the HMF. Specifically, we look at the role stochasticity (or scatter) plays in determining the shape of the QLF. Then, we investigate constraints on feedback processes, AGN/galaxy quenching and local clustering that are set by the amount of scatter in the quasar luminosity versus halo mass relation. Scatter has been a core component surrounding discussions of clustering around quasars based on measurements of luminosity \citep{Shen_2007, Wyithe_2009, Shankar_2010}, and more recently on feedback processes of galaxies in the largest halos \citep{Ren_2019}. In principle, stochasticity in the quasar luminosity to halo mass relation facilitates the probability of an overluminous quasar to reside in a common lower mass halo instead of a very rare, massive halo. This effect is compounded at the massive-end the HMF, because its Schechter-like shape makes lower mass halos exponentially more abundant than the rarest counterparts. Additionally, this also implies that adding scatter in the quasar luminosity versus halo mass relation increases the abundance of the brightest objects. This effect can thus be potentially used to constrain feedback processes if the observed bright end is sufficiently well constrained. The broadening of a distribution due to stochasticity is already well established for a number of observables, e.g. stellar mass and halo mass \citep{Behroozi_2010}, galaxy luminosity and halo mass \citep{Cooray2005, Ren_2019}, black hole mass and stellar mass \citep{Hirschmann_2010}, black hole mass to stellar velocity dispersion \citep{Volonteri_2011}, stellar mass and AGN luminosity \citep{Veale_2014}.

The inclusion of stochasticity in the QLF can be explicitly developed through the conditional luminosity function (CLF) approach that is routinely used in galaxy luminosity modelling \citep{Yang_2003, Cooray2005} just by constructing the median quasar luminosity versus halo mass relation in a way that is consistent with the observed QLF given the quasar duty cycle and the amount of log-normal scatter. The CLF is a powerful tool that links the quasar luminosity directly to its host dark-matter halo, and describes the distribution of luminosities inside a halo of given mass. Qualitatively, we expect scatter to suppress the massive end of the median quasar luminosity to halo mass relation (see \citealt{Ren_2019} for a galaxy analogue).

In this work, we run such analysis for quasars at $z \sim 4$, motivated from the availability of a new robust determination of the LF thanks to the recent survey from the Subaru Hyper Suprime Camera \citep{Akiyama2017}. In addition, this epoch operates as an ideal bridge to make statements on the environments that quasar inhabit at high redshifts ($z > 6$), as well as infer consequences in the galaxy-AGN evolution down to lower redshift.

This paper is structured as follows. In Section~\ref{sec:model} we describe the method used to derive the median quasar luminosity versus halo mass relation given scatter and quasar duty cycle. In  Section~\ref{sec:results} we present the median quasar luminosity versus halo mass relations and we discuss the broader implication of this broadening under physical context of feedback, quenching and local clustering. Section~\ref{sec:conclus} summarizes our findings and concludes. Throughout this paper, we use WMAP-7 cosmology with parameters, $\Omega_{m} = 0.272$, $\Omega_{b} =0.0455$, $\Omega_{\Lambda}=0.728$, $h=0.704$, $\sigma_{8}=0.81$, $n_{s}=0.967$ \citep{2011ApJS..192...18K}. We use the \citet{Jenkins2001} halo mass function. Magnitudes are given in the AB system \citep{1983ApJ...266..713O}.

\section{Modeling}  \label{sec:model}

The typical approach to construct a relation between quasar luminosity and halo mass is through abundance matching which cumulatively matches the number densities of quasar luminosities with those of halos masses \citep{Vale_2004}. This technique is inherently a one-to-one matching process which does not include stochasticity expected to originate from more fundamental processes such as those driving SMBH growth. By taking stochasticity into account, we can expect a qualitatively different picture compared to a deterministic abundance matching. Specifically, the the most luminous quasars are likely to be outliers in terms of accretion rate, i.e. overluminous for their host halo, instead of being hosted within the most massive halos (see Fig.~\ref{fig:sbscalib}). Increasing the magnitude of scatter boosts the probability of the less massive halos to host an overluminous source, as halo abundance is strongly dependent on mass. In fact, compared to the galaxy (UV) luminosity function, it becomes even more critical to consider the effects of stochasticity in the quasar luminosity function where observations typically extends well beyond the $L > L_{*}$ regime. In this work, we largely follow the steps used in \citet{Allen2018} and introduce scatter into our modelling by defining a CLF for quasars. The method is summarized in the following subsections.

\subsection{Distribution of Quasar Luminosities}

We explicitly model the stochasticity with a conditional luminosity function (CLF) approach. Here, the CLF $\Phi(M_{UV}\mid M_{\rm{h}})$ can be interpreted as the probability distribution for quasar magnitudes, $M_{UV}$, given a halo mass $M_{\rm{h}}$,

\begin{equation}
\label{eqn:clf}
\Phi(M_{UV} \mid M_{\rm{h}}) = (1-\epsilon_{\rm{DC}}) \delta(L = 0) + \dfrac{\epsilon_{\rm{DC}}}{\sqrt{2\pi}(2.5 \times \Sigma)}\exp{\Bigg( \dfrac{-\Big[ M_{UV}- M_{UV, \rm{c}}(M_{\rm{h}},\Sigma,\epsilon_{\rm{DC}})\Big]^{2}}{2(2.5 \times \Sigma)^2} \Bigg)}, 
\end{equation}

where $M_{UV,\rm{c}}(M_{\rm{h}})$ is the median quasar magnitude at our given halo mass, $\Sigma$ is the width of the dispersion in dex, $\epsilon_{\rm{DC}}$ is the halo-mass independent quasar duty cycle defined as the constant fraction of SMBHs that have active quasar-mode radiation and $\delta(x)$ is the Dirac delta function. The form of the of the CLF is log-normal and is justified by two reasons: (1) Observationally, the dispersion in $M_{BH} - \sigma$ relation is well fit with a log-normal \citep{Gueltekin2009}. Having $M_{BH}$ as a proxy for quasar luminosity and $\sigma$ to halo mass could plausibly suggest a similar form of intrinsic scatter. (2) The luminosity of a quasar can be considered as product of many processes which can tend to a log-normal dispersion by the central limit theorem. Hence, to first order the scatter in this relation can be thought log-normal. The usual QLF, $\phi(M_{UV})$ can then be derived by the equation,

\begin{equation}
\label{eqn:lf}
\phi(M_{UV}) = \int^{\infty}_{0} \dfrac{dn}{dM_{\rm{h}}} \Phi(M_{UV}\mid M_{\rm{h}}) dM_{\rm{h}}, 
\end{equation}

where $\frac{dn}{dM_{\rm{h}}}$ is the HMF. The median halo mass versus quasar magnitude relation, $M_{UV,\rm{c}}(M_{\rm{h}})$, given $\Sigma$ and $\epsilon_{\rm{DC}}$ can then be obtained by deconvolving Equation~\ref{eqn:lf}. The free parameters in this model are $\epsilon_{\rm{DC}}$, $\Sigma$ and the input QLF, $\phi$. We adopt the $z\sim 4$ QLF data set as reported by \citet{Akiyama2017}. The form of the typical double power law QLF is given by,

\begin{equation}
\label{eqn:olf}
\phi_{\rm{obs}}(M_{UV}) = \dfrac{\phi^{*}}{10^{0.4(\alpha + 1)(M_{UV} - M^{*})} + 10^{0.4(\beta + 1)(M_{UV} - M^{*})}},
\end{equation}

with the \citet{Akiyama2017} parameters we have, $\phi^{*} = 2.66 \times 10^{-7}$ Mpc$^{-3}$ mag$^{-1}$ is the LF normalization factor, $M^{*} = -25.36$ as the characteristic break magnitude, $\alpha = -1.3$ as the faint end slope and $\beta = -3.11$ for the bright end slope. We apply two different deconvolution methods outlined in \citet{Allen2018} and \citet{Ren_2019} to determine $M_{UV,\rm{c}}(M_{\rm{h}})$ through a least-squares fit.

\subsection{Deconvolution Method} \label{sec:deconv}

To prepare Fig~\ref{fig:sbscalib} we follow the deconvolution method outlined in \citet{Allen2018}, further discussed in \citet{Behroozi_2010}. The steps for this iterative process can be summarized as follows:

\begin{enumerate}
  \item We derive the QLF using Equation~\ref{eqn:lf}, given input $\Sigma$ and $\epsilon_{\rm{DC}}$, and assuming the \citet{Jenkins2001} HMF. We start from an initial guess $M_{UV,\rm{c}}' = M_{UV,\rm{c}}(M_{\rm{h}}, \Sigma = 0, \epsilon_{\rm{DC}})$, which can be derived from direct abundance matching, and construct $\phi_{M}(M_{UV}')$ from Equation~\ref{eqn:lf}.
  \item We apply abundance matching between $\phi_{M}(M_{UV}')$ and the input calibration QLF to derive a correction to the relation between $M_{UV}'$ and $M_{UV}$. 
  \item We transform our median quasar magnitude versus halo mass relation, $M_{UV,\rm{c}}'(M_{\rm{h}}, \Sigma, \epsilon_{\rm{DC}})$ according to the relation derived in Step $(2)$. 
  \item Steps $(1)$ to $(3)$ are iteratively repeated. 
  \item The iteration process is terminated when the squared residual difference between successive iteration steps at a fixed number density, $\phi = 6\times10^{-11}$Mpc$^{-3}$ first changes by $(\Delta M_{UV})^2 \leq 0.01$. This choice of the number density corresponds to the value of the fit in the brightest available magnitude data point of \citet{Akiyama2017}.
\end{enumerate}

This deconvolution process approximately derives $M_{UV,\rm{c}}(M_{\rm{h}})$ that best matches the initial calibration QLF in Eq.~\ref{eqn:lf} through progressive improvements, with the caveat that numerical instabilities may arise for a large number of iterations.

\section{Results and Discussion} \label{sec:results}

\subsection{Impact of scatter on the bright end of the QLF}

Fig~\ref{fig:sbscalib} shows one Monte Carlo realization for the distribution of quasar luminosities after sampling from the CLF and HMF, under different values of $\Sigma$. The number of sampled quasar points corresponds to an equivalent cosmological volume of $(\sim 1.5$Gpc$)^{3}$. The figure clearly demonstrates that our treatment of the median halo mass versus quasar magnitude ($M_{UV, {\rm c}}$) relation, successfully returns the $z\sim 4$ QLF (within errors) after assuming values of $\Sigma = 0, 0.3, 0.5$. We include choices for the quasar duty cycle at values of $\epsilon_{\rm{DC}} = 0.01, 1$.

In our modeling where $\Sigma = 0$, i.e. using the typical deterministic abundance matching approach, the resulting function $M_{UV,\rm{c}}(M_{\rm{h}},\Sigma = 0)$ becomes increasingly steep at high halo masses. This is due to the need to abundance-match the exponential drop-off of the HMF at high masses with the power law bright-end of the QLF, and implies that hosts are required to be increasingly efficient at producing quasar luminosity at higher halo masses. We show this behavior in the green curve of Fig~\ref{fig:ratio}, highlighting a positive slope in the ratio of the median quasar luminosity to halo mass. In fact, a similar feature is also present for $\Sigma=0.3$ (red curve, Fig~\ref{fig:ratio}), also yielding a positive slope, albeit the trend is less extreme. We note that the small fluctuations in this curve is just a consequence of the numerical deconvolution technique. For low values of $\Sigma$, the brightest quasars are generally hosted by the most massive halos. In this context, the ratio of median quasar luminosity per unit halo mass over halo mass still shows positive. Identifying a physical mechanism for such a requirement can be difficult in the presence of feedback processes which are regulating SMBH (and galaxy) growth at the high mass end in particular. 

However, Fig~\ref{fig:sbscalib} highlights that the shape of $M_{UV,\rm{c}}(M_{\rm{h}})$ changes when larger $\Sigma$ values are considered. In the case of $\Sigma=0.5$, the median halo mass for a luminous quasar is reduced. This effect is compounded by the exponential increase of the number density of less massive halos. Therefore, the more typical halos around the characteristic halo mass value ($M_{\rm{h}}\lesssim 10^{12.4} M_{\odot}$) begin to play a larger role in shaping the bright end of the QLF for sufficiently large $\Sigma$ ($\Sigma=0.5$ panel of Fig.~\ref{fig:sbscalib}). As a result, the high-mass end of $M_{UV,\rm{c}}(M_{\rm{h}},\Sigma = 0.5)$ flattens out in order for the CLF to successfully fit the observed QLF. Under these circumstances, the shape of the resulting $M_{UV,\rm{c}}(M_{\rm{h}})$ naturally resembles the functional shape of the galaxy luminosity versus halo mass relation (e.g. see \citealt{Ren_2019} Fig.~1), and emphasises a deeper fundamental connection between the galaxies and quasars. In this instance, $M_{UV,\rm{c}}(M_{\rm{h}})$ no longer implies the existence of an increasingly efficient mechanism for quasar accretion as the halo mass grows. Two conclusions can be drawn from the $\Sigma=0.5$ panel: (1) the growth of BHs in the most massive halos is suppressed, and (2) the model suggests that the bright-end of the QLF is populated by objects that are outliers with very high accretion rate compared to the average value for their host halo mass. 

Furthermore, the flattening induced in $M_{UV,\rm{c}}(M_{\rm{h}})$ by large $\Sigma$ values effectively implies that the bright-end of the QLF becomes insensitive to changes in the value of the quasar duty cycle at high halo masses, as the objects are increasingly hosted by relatively common lower-mass halos. This is a generic feature of high $\Sigma$ models and is independent of $\epsilon_{\rm{DC}}$. Thus, models with high values of $\Sigma$ are justified in assuming a mass-independent duty cycle to describe the bright end of the QLF. The weak dependence on halo mass for $\epsilon_{\rm{DC}}$ has been a key feature for several models in describing luminosities of quasars (e.g \citealt{Shankar_2010, Conroy_2012}).

The nature of $\Sigma$ is not restricted to the intrinsic stochasticity in the fundamental processes that build the quasar itself. The dependence on environment also adds an effective `scatter' that is encapsulated in the $\Sigma$ parameter. Thus a high value of $\Sigma$ in $M_{UV,\rm{c}}(M_{\rm{h}})$ could also indicate that powering a quasar accretion mode is contingent on having ideal environmental conditions that funnel sufficient amount of gas to the center of the host galaxy and thus promote SMBH growth. Furthermore, the natural variation in accretion properties as a function of environment is expected to be more considerable for a single AGN compared to variation of luminosity in galaxies, where scatter in star formation efficiency from individual star forming regions is minimized from summing over a population of both star clusters and stars. The significant dependence of quasar properties on the local environment is consistent with results of high redshift ($z>7$) hydrodynamical simulations. For example, \citet{Di_Matteo_2017} identify the conditions of the local tidal field as instrumental to early black hole growth relative to the large-scale overdensity of the host halo.

\subsection{Halo mass threshold for feedback} \label{sec:feedback}

As quasars in less massive hosts contribute significantly to the brightest end of the QLF for sufficient $\Sigma$, under these conditions and to zeroth order, the QLF becomes insensitive to the exact shape of the high-end for both the HMF and  $M_{UV,\rm{c}}(M_{\rm{h}})$ (see the galaxy analogue in \citealt{Ren_2019}). From this, we can approximate the numerical deconvolution process to derive $M_{UV,\rm{c}}(M_{\rm{h}},\Sigma > 0)$ by scaling the $M_{UV,\rm{c}}(M_{\rm{h}},\Sigma = 0)$ relation by a constant factor and by concurrently substituting the high-mass end past some characteristic halo mass  $M^{\rm{c}}_{\rm{h}}$ with a constant luminosity value, i.e.  $M_{UV,\rm{c}}(M_{\rm{h}} > M^{\rm{c}}_{\rm{h}}) = M_{UV,\rm{c}}(M^{\rm{c}}_{\rm{h}})$. We find the best values for the scaling factor and $M^{\rm{c}}_{\rm{h}}$ by least-squares fits based on observed data points $M < M^{*}$. The differences between the deconvolution method described in \citet{Allen2018} and this approximate method are marginal in terms of the resulting quasar luminosity function (a detailed analysis of the differences between these two calculations are highlighted in \citealt{Ren_2019} for galaxies). However, one benefit from the approximate deconvolution method is that $M^{\rm{c}}_{\rm{h}}$ is a well-defined parameter that can be interpreted as a pseudo-scale for feedback.

In Fig~\ref{fig:complf}, we show the extent of our modeled QLFs with this simple approximation method over a large range of scatter, $0.3 < \Sigma < 0.7$ and look at two cases, using as inputs either the \citet{Akiyama2017} or \citet{Kulkarni2019} fits for the $z\sim 4$ QLF. The modeled QLFs are broadly consistent with the data points compiled by the observations for all values of $\Sigma$ considered here. In Fig~\ref{fig:sigdc}, we determine the characteristic feedback threshold $M^{\rm{c}}_{\rm{h}}$, in each of these cases given $\epsilon_{\rm{DC}}$ and $\Sigma$ as fixed parameters. We note that both $\Sigma$ and $\epsilon_{\rm{DC}}$ are fully degenerate with each other for the purpose of determining $M^{\rm{c}}_{\rm{h}}$. Increasing $\Sigma$ or decreasing $\epsilon_{\rm{DC}}$ both lead to a lower $M^{\rm{c}}_{\rm{h}}$. Still, while it is challenging to disentangle the two parameters, a characterization of the inherently constrained parameter space of $\epsilon_{\rm{DC}}$ reveals useful information on the association between halos, galaxies and their AGNs. 

It is evident from Fig~\ref{fig:sigdc} that $M^{\rm{c}}_{\rm{h}}$ depends on the choice of the calibration QLF, and specifically on the robust determination of the QLF for the population around the characteristic $M \sim M^{*}$ quasars. We investigate the magnitude of this effect and compare the distribution of $M^{\rm{c}}_{\rm{h}}$ thresholds across both cases. Quantitatively, the comparison of the two panels shows a small-to-modest difference, $\mathrm{Max } (\Delta\log M^{\rm{c}}_{\rm{h}}) \sim 0.3$, in the distribution of $M^{\rm{c}}_{\rm{h}}$ at fixed $\Sigma$ and $\epsilon_{\rm{DC}}$, with higher $\Sigma$ and lower $\epsilon_{\rm{DC}}$ values having the largest variation in $M^{\rm{c}}_{\rm{h}}$ between our choices of the calibration QLF. 

A point of interest for this figure is that a number of studies have noted luminous quasars to preferentially reside in $\sim 10^{12} M_{\odot}$ hosts, coinciding with the halo mass range that has maximal specific star formation efficiency (e.g. \citealt{Conroy_2012} and references therein). This idea is appealing as it would suggest a single origin for the joint regulation of the galaxy and black hole growth. However in our model, $M^{\rm{c}}_{\rm{h}} \sim 10^{12} M_{\odot}$ requires $\epsilon_{\rm{DC}} \lesssim 10^{-2}$ for either choices of the QLF. Thus, if larger duty cycles are present at high $z$, the two processes would appear to be distinct and affected by feedback operating at different halo-mass scales. Furthermore, the situation can be more complex as we have assumed a monotonic relation between halo and quasar luminosity. In fact the numerical deconvolution method of \citet{Behroozi_2010} is limited to finding the solution under this condition. It is also entirely possible that $M_{UV,\rm{c}}(M_{\rm{h}})$ experiences a turnover after SMBH growth becomes self-regulated at the highest masses. That would have minimal impact on the QLF for high $\Sigma$ values as those rare sources in high-mass halos would be mixed with the population of more common halos, but it would make it increasingly difficult reconcile a self-consistent single halo-mass scale for quasar and galaxy feedback.

Deriving constraints on the average duty cycle $\epsilon_{\rm{DC}}$ at high redshifts remain an open problem, as it critically depends on a variety of unsolved processes, including but not limited to the formation of seeds and the dominant mode of early SMBH growth. Clustering studies provide one avenue of partially constraining $\epsilon_{\rm{DC}}$,  however results are mixed due to the wide range of environments occupied by quasars, hence resulting in duty cycles between $10^{-3} \gtrsim \epsilon_{\rm{DC}} \gtrsim 6\times10^{-1}$ for $z \sim 4$ quasars \citep{Shen_2007,He_2017}. On the other hand, \citet{Aversa_2015} infer an average $\epsilon_{\rm{DC}} \sim 1$ at $z > 3$ using a physically motivated light-curve parameterisation to derive the BH mass function from the AGN luminosity function, which corresponds to a mass threshold $M^{\rm{c}}_{\rm{h}} > 10^{12.9} M_{\odot}$. Additionally, we can draw a comparison using $M_{UV,\rm{c}}(M_{\rm{h}})$ from the hydrodynamical suite MassiveBlack-II (MBII, \citealt{Khandai_2015}) with $\epsilon_{\rm{DC}} \sim 1$, finds a scatter of $\Sigma \sim 0.55$ for $M_{\rm{h}} \sim 10^{11.8} M_{\odot}$ halos. The MBII simulation volume of $(100\mathrm{Mpc}/\mathrm{h})^{3}$ is capable of hosting halos up to $M_{\rm{h}} \sim 10^{12.7} M_{\odot}$, but only expects $\sim 100$ halos with $M > 10^{12} M_{\odot}$. Thus, despite an insufficient simulation volume to fully capture the flattening in $M_{UV,\rm{c}}(M_{\rm{h}})$, the MBII analysis is consistent with a halo mass threshold of $10^{12.9} M_{\odot} \lesssim M^{\rm{c}}_{\rm{h}} \lesssim 10^{13.2} M_{\odot}$ (see Appendix~\ref{apdx:a} for further details on modeling consistency with MB-II). From a physical interpretation perspective, assuming that AGN feedback can act independently in quenching galaxy and SMBH growth \citep{Cielo_2018} would naturally lead to a separation of the characteristic halo masses where star formation and black hole growth become affected.

\subsection{Consequences on luminosity-based clustering measurements}
Rare quasars detected at the epoch of cosmic dawn are understood to be powered by SMBH of masses around $10^{9} M_{\odot}$, hence there is a general expectation that these quasars trace extreme overdensities and reside within massive hosts. However, attempts at observational confirmation based on quasar clustering have historically reported mixed results, in particular at high redshift when quasar counts are increasingly sparse. For example, at $z\sim 4$ high biases have been reported using quasar-quasar correlation function measurements \citep{Shen_2007, Onoue_2017}, while other studies have found little to no evidence of significant clustering based on quasar-galaxy cross correlation \citep{Fukugita_2004, He_2017}. Likewise, a number of studies (both simulations and observations) around $z\sim 5-6$ quasars have shown that they belonged to a wide range of environments: overdensities \citep{Stiavelli05, RomanoDiaz2011, Husband_2013, Costa_2014, Morselli_2014, Garc_a_Vergara_2017} or average/underdensities \citep{Kim_2009, Mazzucchelli_2017, Champagne_2018, Ota_2018}. Compared to bright galaxy samples, that have been widely used to infer luminosity dependence of galaxy-halo properties (see \citealt{Zheng_2009, Trenti_2012, Harikane_2017}), luminous quasars are rarer. Hence, small number count stochasticity intrinsically limits the ability to draw robust inference from current observations. 

From Fig.~\ref{fig:sbscalib}, it is evident that both $\epsilon_{\rm{DC}}$ and $\Sigma$ impact local clustering by reducing the average mass of the quasar host. To demonstrate this explicitly, we derive the distribution of the linear bias factor, $b$ for our quasar host halos using:

\begin{equation}
\label{eqn:rclf}
p(b \mid M_{UV}) =  \dfrac{\dfrac{dn}{dM_{\rm{h}}}(M_{\rm{h}}(b)) \Phi(M_{\rm{h}} \mid M_{UV})}{\phi(M_{UV})}
\end{equation}

where $M_{\rm{h}}(b)$ is taken here as the inversion of the analytical \citet{Sheth_1999} bias relation, and $\Phi(M_{\rm{h}} \mid M_{UV})$ is the inverse conditional luminosity function for the distribution of halo masses given some quasar magnitude, $M_{UV}$. In Fig.~\ref{fig:bias} we show the range of linear bias for quasars in 2 groups: quasars that populate the extreme bright end ($M_{UV} = -28$) and the faint end ($M_{UV} = -22$). The sharp peak feature is a consequence of imposing a flat cutoff in $M_{UV,\rm{c}}(M_{\rm{h}})$ and we do not expect this to impact the results in any significant qualitative way. It is clear that $\Sigma$ and $\epsilon_{\rm{DC}}$ dictate the variety of environments for quasars more luminous than the characteristic magnitude. For the most luminous quasars, the dispersion in the bias is predominantly dependent on the threshold value, $M^{\rm{c}}_{\rm{h}}$ and the spread in $M_{\rm{h}}$ at a quasar magnitude to a lesser extent. This is because the bias factor has a strong non-linear dependence in $M_{\rm{h}}$. The non-linearity also induces a positive skew that scales with $\Sigma$ in the bias distribution, suggesting an uneven distribution of environments around the mode value of the bias. In contrast, the distribution of the linear bias around fainter quasars is seen to be substantially insensitive to $\Sigma$, but not to $\epsilon_{\rm{DC}}$. This highlights the opportunity to use the local clustering around fainter quasars as a probe to constrain the quasar duty cycle. Indeed, such a task is easily within reach using next-generation facilities such as the James Webb Space Telescope, which will be able to probe with both imaging and spectroscopy the fainter companions around high-$z$ quasar halos. One caveat to note is that Fig~\ref{fig:sbscalib} has been obtained using the median quasar magnitude versus halo mass relation inferred from the \citet{Akiyama2017} QLF. Using the relation derived from the \citet{Kulkarni2019} QLF would both shift to more positive values and broaden the linear bias distribution. Therefore, the analysis is dependent upon a precise determination of the QLF.  

In addition, it is important to stress that large values of $\Sigma$ dilute signatures of luminosity dependent clustering, assuming a quasar duty cycle largely independent of halo mass. We check using the high/low luminosity bins of \citet{He_2017}, taking halo biases for quasars of $M_{UV} \sim -25.5$ and $M_{UV} \sim -23.5$ we find that our entire range of $\Sigma > 0.3$ models is consistent with no luminosity dependent clustering between the two bins. This is generally in agreement with the conclusion of \cite{He_2017}, however we still find a weak luminosity dependence for clustering if we extend the baseline of our luminosity bins. In fact, we predict that the bias of the brightest quasars ($M_{UV} \sim -28$) can still be quantified as different from that of the population of faint quasars ($M_{UV} \sim -22$) to a confidence of $97.5\%$ provided that $\Sigma \sim 0.5$.

Future wide-area surveys are required to create a representative sample of brighter $M_{UV} < -25.5$ quasars in order to conclusively establish the luminosity dependence of quasar clustering.

\section{General Remarks \& Conclusion} \label{sec:conclus}

\begin{figure}[h!]
	\centerline{\includegraphics[angle=-00, scale=0.60]{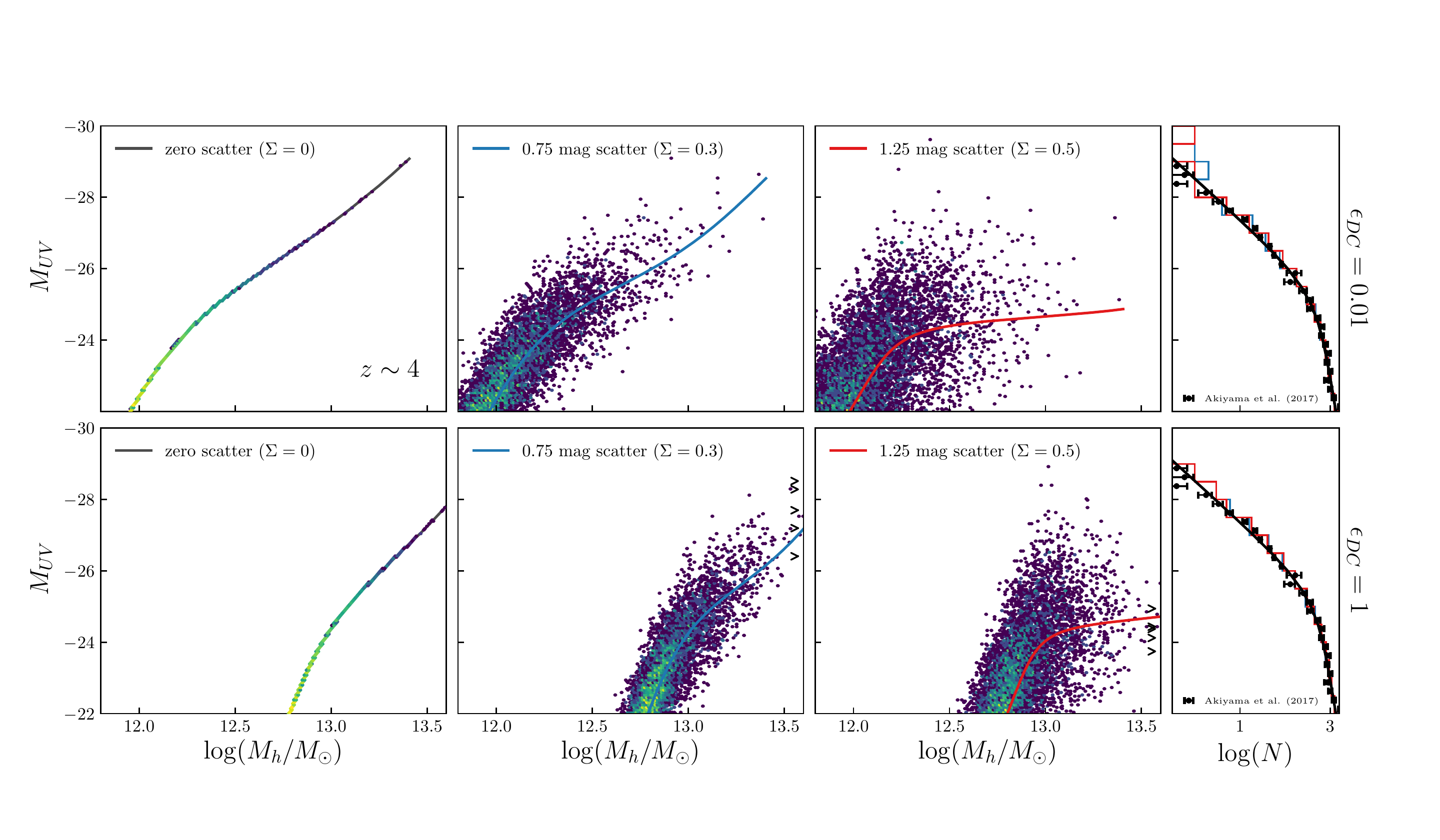}}
	\caption{\small Left-most three columns of panels: Simulated $z=4$ quasar magnitudes as a function of host halo mass as derived from our conditional luminosity function model, assuming a $(1.5\mathrm{Gpc})^3$ comoving volume, and different values for log-normal scatter $\Sigma$ and quasar duty cycle $\epsilon_{\rm{DC}}$ in each panel. The median of the data points is shown as solid colored line. Arrows on the right edge of the bottom row of plots show quasars with the indicated luminosity inside halos beyond plot limits. The rightmost column has panels showing the resulting quasar luminosity functions from binning the model points as histograms (colored), compared to the input calibration function (solid black).}
	\label{fig:sbscalib}
\end{figure}

\begin{figure}[h!]
	\centerline{\includegraphics[angle=-00, scale=0.60]{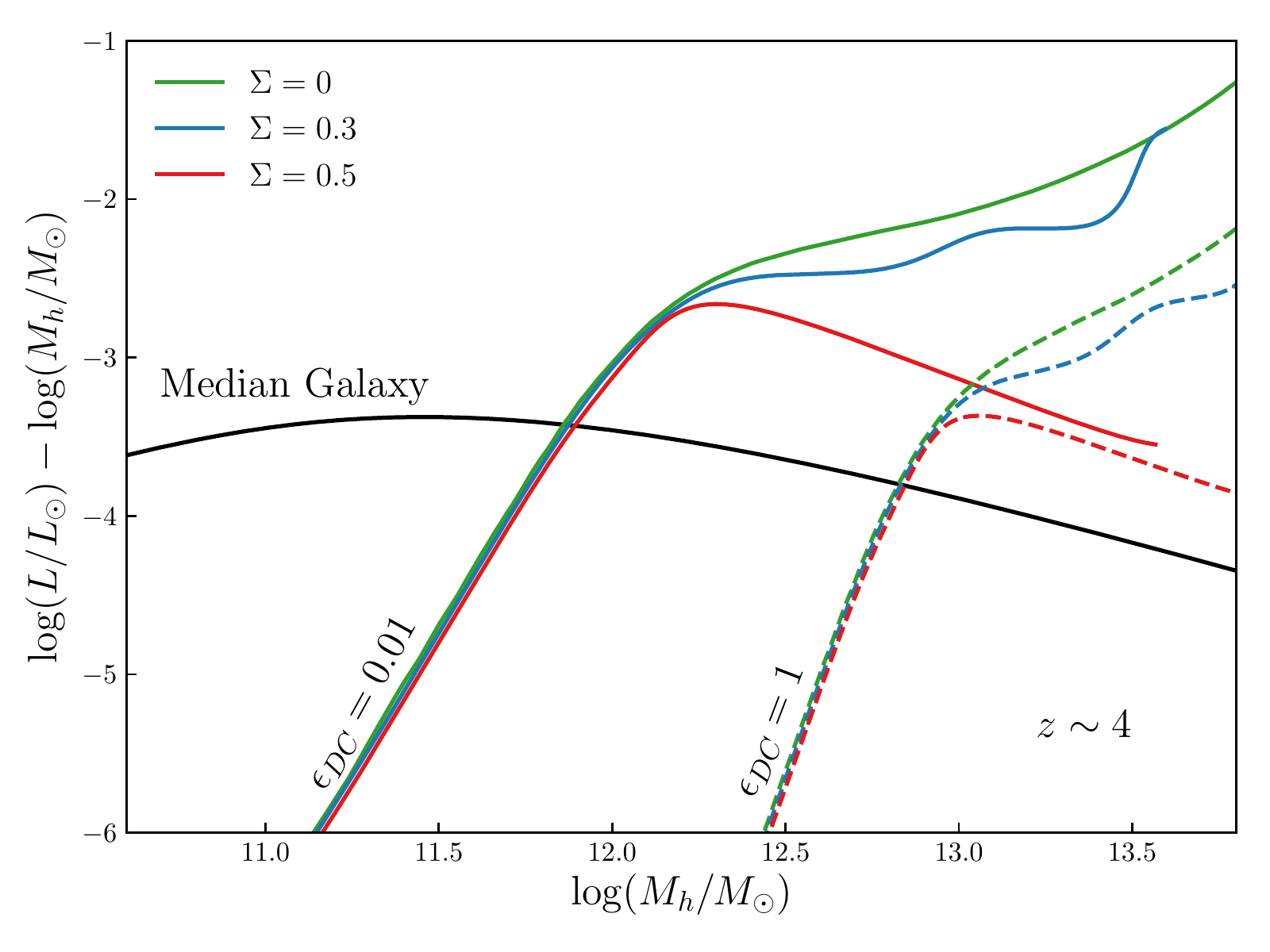}}
	\caption{\small Quasar luminosity to halo mass ratio as a function of halo mass, $\log(\frac{L/L_{\odot}}{M_{\rm{h}}/M_{\odot}})$ for different values of scatter $\Sigma$ (colored lines) and duty cycle $\epsilon_{\rm{DC}}$ (solid vs dashed lines) at $z\sim4$. For comparison, the same relation is also shown for galaxies (black solid line, assuming no scatter and unity duty cycle). The oscillations in the curves for $\Sigma > 0$ are from numerical instabilities in the \citet{Allen2018} deconvolution process.}
	\label{fig:ratio}
\end{figure}

\begin{figure}[h!]
	\centerline{\includegraphics[angle=-00, scale=0.60]{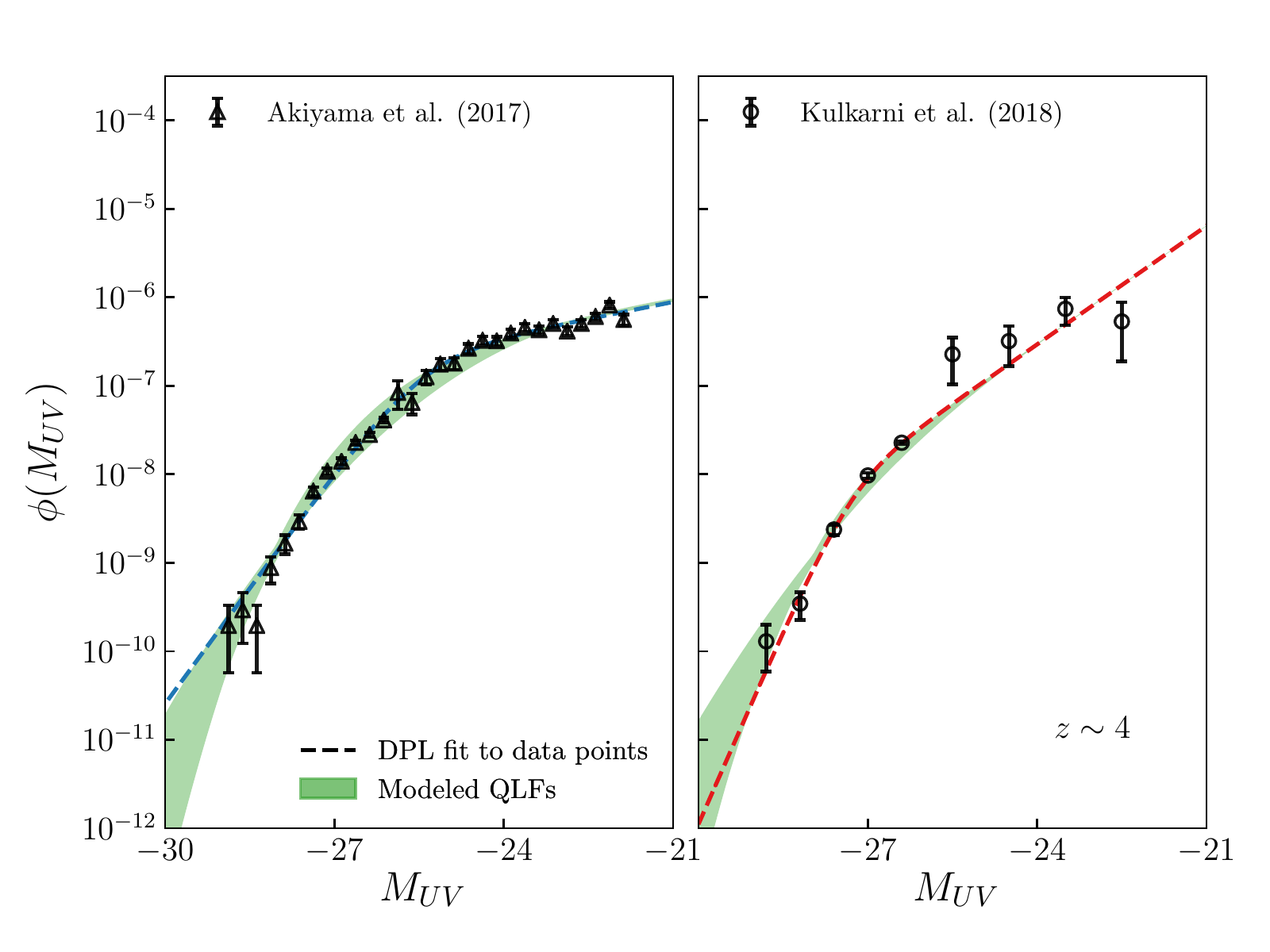}}
	\caption{\small Modeled QLFs at $z \sim 4$ for a range of scatter, $0.3 < \Sigma < 0.7$ (shaded area), assuming different observed QLFs as inputs (black data points). The left panel corresponds to an input using \citet{Akiyama2017} QLF and the right panel is for the \citet{Kulkarni2019} QLF. The dashed lines are the respective best fit double power law to the data points (see Eq.~\ref{eqn:olf}). The parameters for the fits are: (Left panel, \citealt{Akiyama2017}) $\log(\phi^{*}) = -6.58$, $M^{*} = -25.36$, $\alpha = -1.3$ and $\beta = 3.11$; (Right panel, \citealt{Kulkarni2019}) $\log(\phi^{*}) = -7.99$, $M^{*} = -27.28$, $\alpha = -2.11$ and $\beta = -4.64$.}
	\label{fig:complf}
\end{figure}

\begin{figure}[h!]
	\centerline{\includegraphics[angle=-00, scale=0.60]{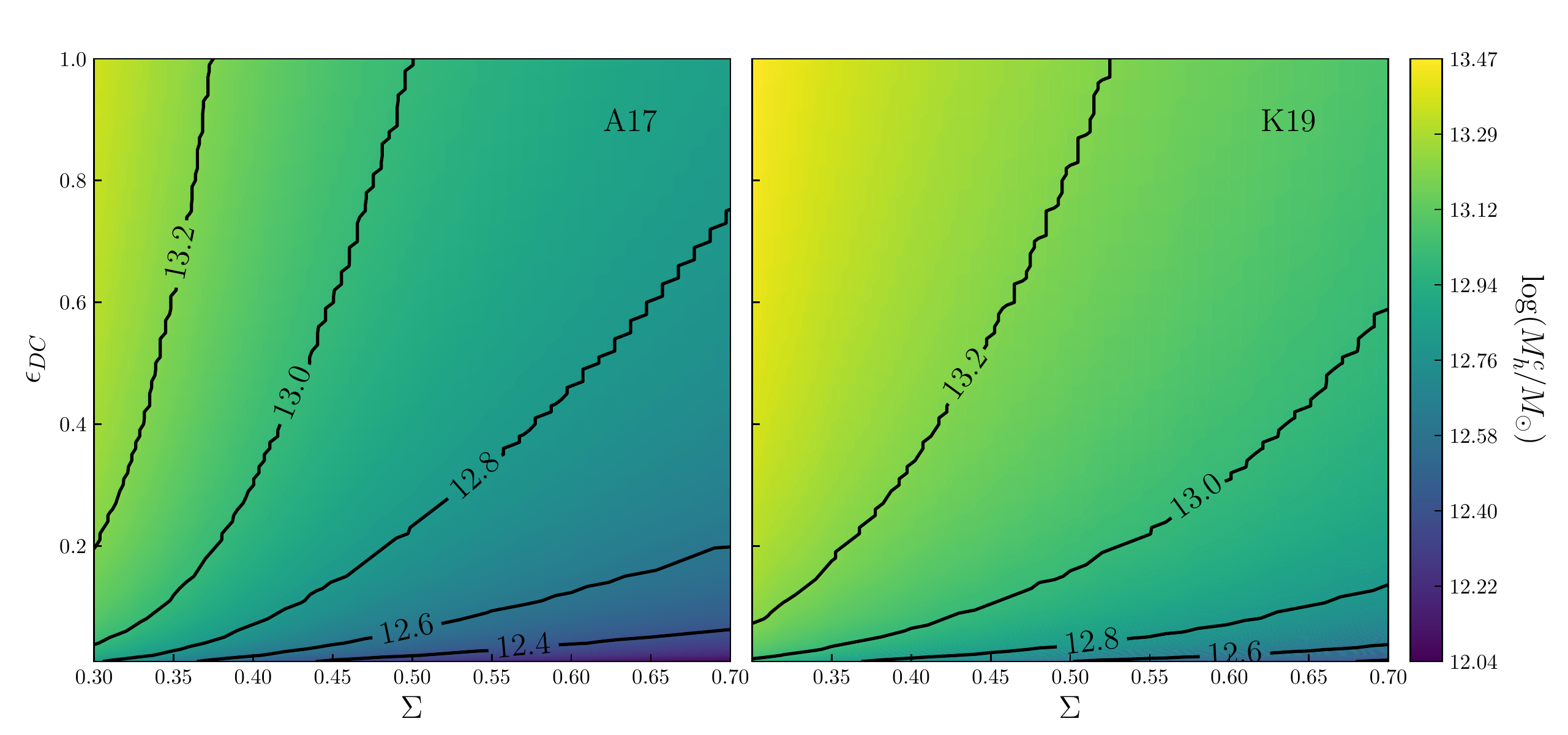}}
	\caption{\small Distribution of values for the critical halo mass threshold $M_{\rm{h}}^c$ as a function of input scatter $\Sigma$ and duty cycle $\epsilon_{\rm{DC}}$. The left panel corresponds to models using the \citet{Akiyama2017} QLF, while the right panel is for the \citet{Kulkarni2019} QLF. The solid black lines are contours for select characteristic halo masses ($M_{\odot}$ log-scale).}
	\label{fig:sigdc}
\end{figure}

\begin{figure}[h!]
	\centerline{\includegraphics[angle=-00, scale=0.80]{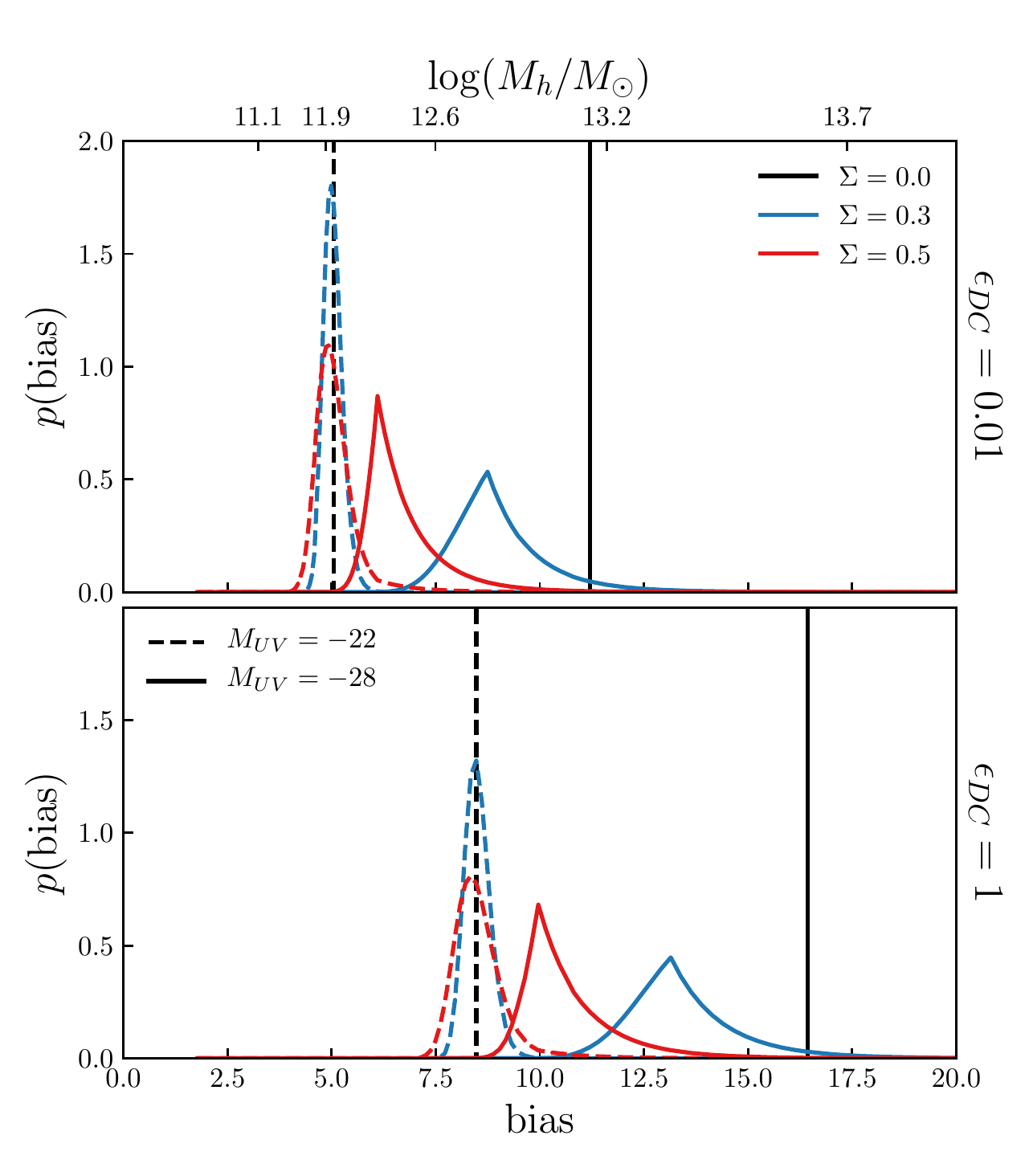}}
	\caption{\small Probability distribution of the quasar bias value for different model input parameters $\Sigma$ (colored lines) and $\epsilon_{\rm{DC}}$ (top vs bottom panel). The halo mass corresponding to the bias value on the horizontal axis is shown on top of the upper panel. These distributions have been derived using the median quasar magnitude versus halo mass relations in our model with the \citet{Akiyama2017} QLF as inputs.}
	\label{fig:bias}
\end{figure}

The broad power-law bright end of the quasar luminosity function (QLF) relative to the exponential drop-off in the host halo mass function (HMF) suggests there could be significant stochasticity, $\Sigma$ in the quasar magnitude versus halo mass relation, $M_{UV,\rm{c}}(M_{\rm{h}})$. In this work, we use a conditional luminosity function approach to derive $M_{UV,\rm{c}}(M_{\rm{h}})$ from the observed $z \sim 4$ QLF assuming values for our free parameters scatter $\Sigma$ and the quasar duty cycle $\epsilon_{\rm{DC}}$, and investigate how these parameters shape $M_{UV,\rm{c}}(M_{\rm{h}})$. In addition to a full deconvolution study, we also construct an approximate best-fit $M_{UV,\rm{c}}(M_{\rm{h}})$, with a functional form characterized by a constant value above a critical halo mass threshold at $M^{\rm{c}}_{\rm{h}}$, i.e. $M_{UV,\rm{c}}(M_{\rm{h}} > M^{\rm{c}}_{\rm{h}}) = M_{UV,\rm{c}}(M^{\rm{c}}_{\rm{h}})$. In this framework, $M^{\rm{c}}_{\rm{h}}$ can be interpreted as the critical value beyond which feedback is required to significantly regulate black-hole growth/quasar radiation to avoid over-producing luminous objects that are not observed from the QLF. Finally, we investigated how this threshold depends on model assumptions and parameters, discussing physical interpretations of our model, with the following key results:  

\begin{itemize}

\item We show that $\Sigma$ induces a flattening in $M_{UV,\rm{c}}(M_{\rm{h}})$ to account for the abundance of lower mass quasar hosts with extreme accretion rates populating the bright end of the QLF (Fig.~\ref{fig:sbscalib}). The flattening effect from stochasticity has been previously explored for a variety of relations (see examples: stellar mass, \citealt{Behroozi_2010}; galaxy luminosity, \citealt{Ren_2019}; BH mass/AGN luminosity, \citealt{Aversa_2015}). We find that values of $\Sigma < 0.3$ lead to a rising quasar magnitude-halo mass ratio for more massive halos. This can be difficult to reconcile with ideas of self-regulating black hole growth. In contrast, $\Sigma \gtrsim 0.5$ implies a flattened $M_{UV,\rm{c}}(M_{\rm{h}})$ for massive halos, and is indicative of a turnover in quasar efficiency. This suggests both that the median black hole growth is regulated at the massive halo end and that there is an increasing likelihood that the most luminous quasars are extreme outliers in accretion efficiency hosted in relatively common medium-mass dark-matter halos. 

\item Following this, we note that constant duty cycle is a good approximation for modeling the bright end of the QLF assuming significant scatter ($\Sigma\gtrsim 0.3$). Since halo abundance is strongly dependent on mass, the abundance of lower mass halos with extreme accretion rates dominates over the influence of any changes in duty cycle for larger quasar hosts. 

\item The characteristic mass threshold for feedback (defined as the halo mass, $M^{\rm{c}}_{\rm{h}}$ where $M_{UV,\rm{c}}(M_{\rm{h}})$ flattens) is strongly dependent on $\Sigma$ and $\epsilon_{\rm{DC}}$ (Fig.~\ref{fig:sigdc}). $M^{\rm{c}}_{\rm{h}}$ is relatively insensitive to variations in observed QLF determinations, with Max($\Delta\log M^{\rm{c}}_{\rm{h}}) \sim 0.3$ between the use of the \citet{Akiyama2017} and of the \citet{Kulkarni2019} QLF as modeling inputs.

\item $\Sigma$ and $\epsilon_{\rm{DC}}$ are strongly degenerate. An increase in $\Sigma$ has essentially the same impact in decreasing $M^{\rm{c}}_{\rm{h}}$ as a reduction in $\epsilon_{\rm{DC}}$. Disentangling this degeneracy would require additional measurements, such as local clustering using cross-correlations between quasars and galaxies around faint ($M_{UV}\sim -22$) quasars (Fig.~\ref{fig:bias}). These complementary observations would need to reach about a factor $10\times$ fainter than the quasar ($-20 \lesssim M_{UV}\lesssim -19$). This limit is already within imaging capabilities of current facilities (e.g., Wide Field Camera 3 instrument on the Hubble Space Telescope), and sufficiently bright for spectroscopy with the upcoming James Webb Space Telescope.

\item Matching the halo mass for optimal efficiency in both quasar radiation emission and stellar formation in galaxies, $\sim 10^{12} M_{\odot}$, requires $\epsilon_{\rm{DC}} < 10^{-2}$. Observations of clustering and hydrodynamical simulations infer a range from $10^{-3} < \epsilon_{\rm{DC}} < 1$. A high $\epsilon_{\rm{DC}} \sim 1$ would return $M^{\rm{c}}_{\rm{h}} \gtrsim 10^{12.9} M_{\odot}$ for quasars hosts, which could suggest that AGN feedback quenches galaxies and SMBH growth independently of each other \citep{Cielo_2018}. 

\item We calculate the distribution of halo bias around bright ($M_{UV} = -28$) and faint ($M_{UV} = -22$) quasar hosts (Fig.~\ref{fig:bias}). $\Sigma$ increases the spread of the biases, while a rising $\epsilon_{\rm{DC}}$ increases the median halo bias. $\Sigma > 0.3$ weakens the luminosity dependence for clustering such that there is an effectively no luminosity dependence for clustering \citep{He_2017}.  

\end{itemize}

The framework developed in this manuscript allows us to speculate on the evolution of quasar demographics across $z$. For example, if we make the assumption that the magnitude of scatter, $\Sigma$ and the halo mass threshold, $M^{\rm{c}}_{\rm{h}}$ remain relatively unchanged across redshift, then we would generally expect the bright end slope of the QLF to become shallower at higher $z$ from the decreased abundance of $M > M^{\rm{c}}_{\rm{h}}$ halos. However, the change in $\Sigma$ over $z$ is not well constrained observationally owing to a number of factors, such as the intrinsic rarity of the brightest objects and the challenges in observing the more typical $M \sim M^{*}$ objects. A recent work by \citet{Marshall_2019} investigating the primary growth mechanisms of SMBHs finds that merger-driven SMBH growth is subdominant compared to instability-driven growth at $z > 2$. The smaller contribution of mergers on the mass history of SMBHs can point to $\Sigma$ having a weaker dependence on $z$. The assumption that $M^{\rm{c}}_{\rm{h}}$ is independent over $z$ can be justified on a theoretical basis from the relative $z$ independence in halo-mass where we expect radio-mode feedback to become non-negligible \citep{Croton_2006}. Additionally, the mass-independence from this mode of feedback is also partially supported through sophisticated empirical modeling of stellar-mass to halo-mass ratio (SHMR) demonstrating a weak evolution in the peak of the SHMR relation across redshifts \citep{Tacchella_2018, Behroozi_2019}. 

However, it should be noted that the current extent of available evidence through observations does not yield a compelling case on the direction the bright end slope should evolve with redshift. In one case, the QLFs from \citet{Akiyama2017} and \citet{Matsuoka2018} shows that the bright end slope evolves to be more shallow at higher $z$ (at least from $z > 4$), while empirical modeling by \citet{Kulkarni2019} suggests the contrary, that the bright end slope should steepen towards higher $z$. It is worthwhile to note that the bolometric QLFs recently recompiled by \citet{Shen_2020} suggests for the bright end slope to become shallower at higher $z$ (from $z\sim 2$) which is broadly consistently with the picture provided here. However, one surprising element is that \citet{Shen_2020} finds the bright end slope to also become shallower at $z < 2$. A conclusive observational picture of the evolution in the QLF across a large range of magnitudes is therefore essential in order to answer such questions. This task is both challenging and time consuming, requiring the need to leverage both: (1) wide-field surveys to capture abundance of the rarest and brightest of quasars, and (2) deep imaging plus spectroscopy together with extensive modeling to deconvolve the contribution of quasar light within its host galaxy for the characteristic $M \sim M^{*}$ quasars. In this context, simple but effective models that physically capture the dominant contribution in the evolution of the QLF can provide effective tools to assess expectations in preparation of future surveys.

\acknowledgements
This research was conducted by the Australian Research Council Centre of Excellence for All Sky Astrophysics in 3 Dimensions (ASTRO 3D), through project number CE170100013. K.R is additionally supported through the Research Training Program Scholarship from the Australian Government. TDM acknowledges funding from NSF ACI-1614853, NSF AST-1517593, NSF AST-1616168, NASA ATP 80NSSC18K1015 and NASA ATP 17-0123.

\appendix
\section{Comparison of modelings methods and simulations} \label{apdx:a}

\begin{figure}[h!]
	\centerline{\includegraphics[angle=-00, scale=0.60]{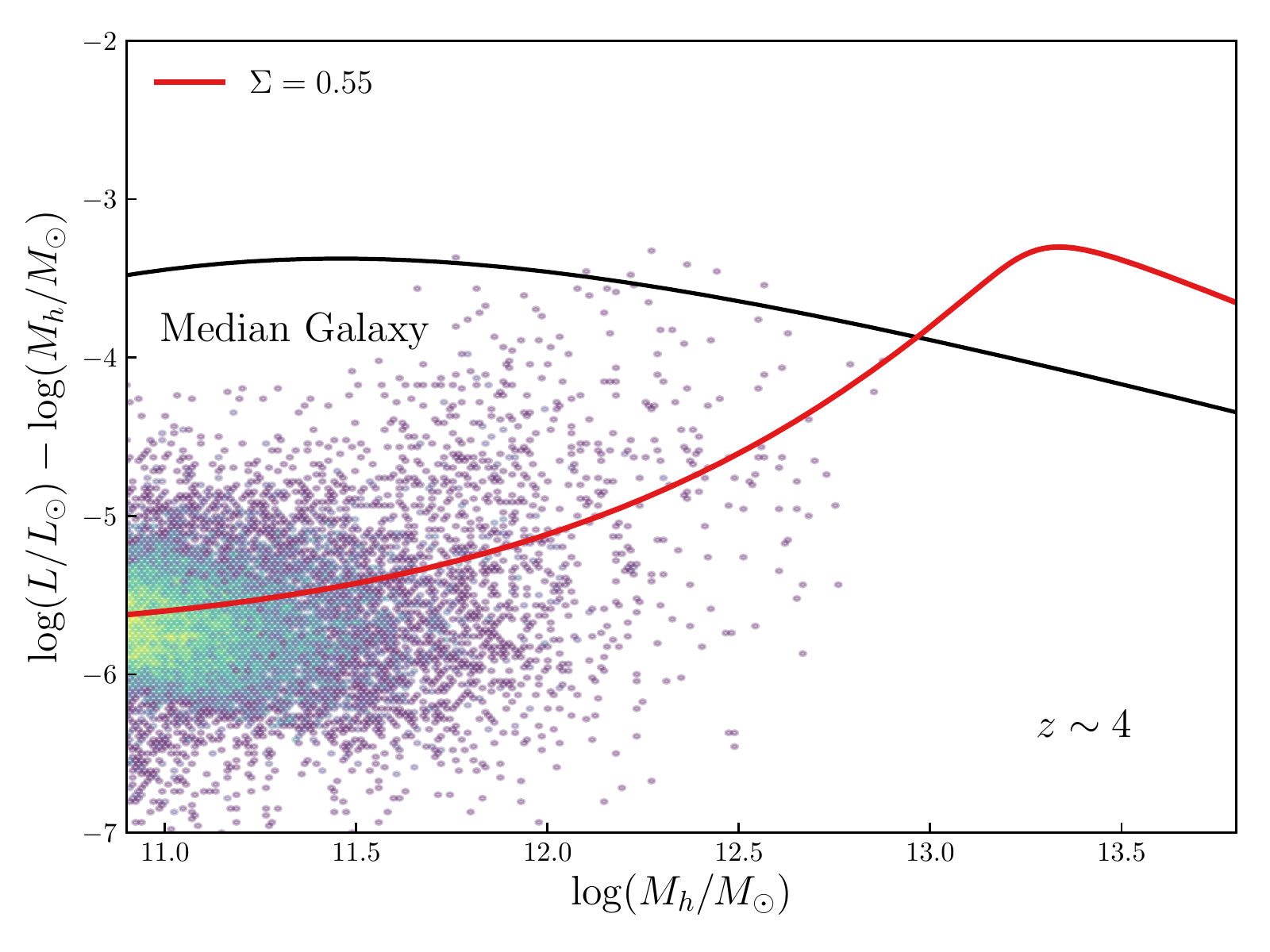}}
	\caption{\small Quasar luminosity to halo mass ratio as a function of halo mass, $\log(\frac{L/L_{\odot}}{M_{\rm{h}}/M_{\odot}})$ for simulated $z = 4$ quasars from MassiveBlack-II. The red solid line represent the output from our modeling for scatter $\Sigma = 0.55$ (corresponding to the scatter in MassiveBlack-II quasars inside $M_{\rm{h}} \sim 10^{11.8} M_{\odot}$ halos) assuming the \citet{Kulkarni2019} QLF as an input and a duty cycle $\epsilon_{\rm{DC}} = 1$ at $z \sim 4$. For comparison, the same relation is also shown for galaxies (black solid line, assuming no scatter and unity duty cycle).}
	\label{fig:kratio}
\end{figure}

In Section~\ref{sec:feedback}, we determine a range for the halo-mass threshold in MassiveBlack-II (MB-II). In Fig~\ref{fig:kratio}, it is clear that MB-II lacks the volume to capture any possible turnover in $M_{\rm{h}}$. The output displayed is generated from the deconvolution process outlined in Section~\ref{sec:deconv}. The input uses $\Sigma = 0.55$ corresponding to the scatter in MB-II quasars inside $M_{\rm{h}} \sim 10^{11.8} M_{\odot}$ halos and assumes the \citet{Kulkarni2019} QLF, which is a closer fit to faint quasars present in the MB-II QLF. The modeling shown here indicates a turnover at $M_{\rm{h}} \sim 10^{13.3} M_{\odot}$, slightly higher than the one inferred through our approximate modeling at $M_{\rm{h}} \sim 10^{13.2} M_{\odot}$. As discussed in Section~\ref{sec:feedback}, the key parameter between the choice of the QLF that impacts $M^{\rm{c}}_{\rm{h}}$ is the position and normalization of the characteristic magnitude break, $M^{*}$, hence only a small difference in $M^{\rm{c}}_{\rm{h}}$ is expected when using either \citet{Akiyama2017} and \citet{Kulkarni2019} QLFs.

\bibliography{qlfscatter}
\bibliographystyle{aasjournal}
\end{document}